\documentclass[a4paper,11pt]{article}
\usepackage{graphicx}
\usepackage{amsmath}

\input{tcilatex}

\begin{document}

\date{\today}

{\large \textbf{On Essential Incompleteness of Hertz's Experiments
on Propagation of Electromagnetic Interactions}}

\bigskip \bigskip

$\;\;\;\;\;\;\;\;\;\;$\textbf{R. Smirnov-Rueda}

\bigskip

$\;\;\;\;\;\;\;\;\;\;$\textit{Applied Mathematics Department, }

\textit{$\;\;\;\;\;\;\;\;\;\;$Faculty of Mathematics }

\textit{$\;\;\;\;\;\;\;\;\;\;$Complutense University, }

\textit{$\;\;\;\;\;\;\;\;\;\;$28040 Madrid, Spain }

\bigskip

\begin{abstract}

The historical background of the 19th century electromagnetic
theory is revisited from the standpoint of the opposition between
alternative approaches in respect to the problem of interactions.
The 19th century electrodynamics became the battle-field of a
paramount importance to test existing conceptions of interactions.
Hertz's experiments were designed to bring a solid experimental
evidence in favor of one of them. The modern scientific method
applied to analyze Hertz's experimental approach as well as the
analysis of his laboratory notes, dairy and private letters show
that Hertz's "\textit{crucial}" experiments cannot be considered
as conclusive at many points as it is generally implied. We found
that alternative Helmholtz's electrodynamics did not contradict
any of Hertz's experimental observations of transverse components
as Maxwell's theory predicted. Moreover, as we now know from
recently published Hertz's dairy and private notes, his first
experimental results indicated clearly on infinite rate of
propagation. Nevertheless, Hertz's experiments provided no further
explicit information on non-local longitudinal components which
were such an essential feature of Helmholtz's theory. Necessary
and sufficient conditions for a decisive choice on the adequate
account of electromagnetic interactions are discussed from the
position of modern scientific method.
\end{abstract}


%

\section{Introduction}

Throughout the history of science, one can trace the importance of
the notion of interaction in contrasting rival physical theories.
Explanation of how remote particles influence each other through
the empty space, challenged inquisitive minds of nearly all
outstanding thinkers of earth civilization. In practice, we always
observe one object acting on another by contact or at a distance.
Before Newton introduced his revolutionary universal gravitation
theory, the realist natural philosophy deriving from Aristotle
considered only contact action. In its way to the practical
knowledge the medieval science were laying down a new scientific
method, breaking old religious prejudices and relying only on the
simplest and clearest propositions. Postulation of undetectable,
hidden or invisible causes responsible for interaction between
material objects was inconceivable.

The obvious break of that consensus came with the Newton's
discovery of universal gravitation as an action-at-a-distance
theory. It won immediate acceptance in England but not on the
Continent, where the Cartesians regarded the notion of newtonian
gravitation as an occult immaterial influence. This controversy
took place at the time of the first scientific revolution when new
conceptions were arising and changing very rapidly, hampered
sometimes by confusions and prejudices. The arguments which lead
medieval scholars to accept one or another theory were not only
empirical or theoretical arguments. They also aroused partly out
of the logical requirements of a particular metaphysics.
\textit{Matter cannot act where it is not} defined the cartesian
metaphysical prohibition against \textit{action-at-a-distance}.
Types of scientific explanation began dependent upon the $model$
of nature, i.e. could have been made precise only in the context
of analogies or presupposed models.

The obvious success of Newton's theory in astronomy, similarity of
electric and magnetic phenomena to gravitation confirmed later by
Coulomb and Amp\`{e}re, and the growing influence of empiricism in
natural philosophy helped to modify the general consensus in favor
of the direct action-at-a-distance. Mathematical models for
hydrodynamics made it possible to describe the mechanism of
transmission of influences through the continuous media. It was
generally assumed that the gravitational attraction was
transmitted instantaneously and, hence, always acted along the
line joining the simultaneous position of two bodies. Bearing in
mind the accuracy-limit of astronomical data at that time,
Laplace$^{(1)}$ in 1799 calculated that the velocity of
propagation of gravitational interaction must have been at least
eight orders of magnitude greater than that of light (modern
discussion of this issue can be found in$^{(2)}$). This
respectable conclusion gave major support to the validity of the
\textit{instantaneous action-at-a-distance (IAAAD)} concept,
leaving open the question of the physical cause of gravity. On
this latter subject, a conceptual arm-wrestle was initiated
between supporters and detractors of IAAAD. Newton himself had
already thought that some physical mediator must exist. \textit{It
is absurd}, he said$^{(3)}$, \textit{to suppose that gravity is
innate and acts without a medium, either material or immaterial}.
Many eminent scientists like Laplace explained the phenomenon as
due to an \textit{impulsion} of some immaterial fluid, but did not
find it appropriate at the time to search for a reliable physical
explanation of the cause of interaction, and so they bequeathed
this task to future generations. \textit{There is no need at all},
Laplace$^{(4)}$ declared in 1796, \textit{to posit vague causes,
impossible to submit to analysis, and which the imagination
modifies to its liking in order to explain these phenomena}.

At that stage, it is not surprising that for many of the
adversaries of IAAAD these suggestions of \textit{immateriality}
could not be dissociated from spiritual, religious and other
non-scientific notions. This path of reasoning led them to
conclude that if the IAAAD concept did not imply any material
mediator, then it did not imply any physical mediator at all.
Thus, in their opinion IAAAD became inconceivable from the point
of view of common-sense logic. This prepared the ground for the
reappearance of an alternative and half-forgotten Aristotelian
concept, namely, that of action by local contact (or contact
action)$^{(5)}$. In its modern form known as the local field
concept, it was reintroduced by Faraday, giving apparently a more
satisfactory physical description of the interaction phenomena. It
was based on the causality of local interactions in terms of
material forces filling space. After the discovery of
electromagnetic induction, Faraday became convinced that electric
and magnetic influences were propagated through a material medium
rather than at a distance through an empty space.

Thus, nearly one and a half century after Newton's discovery of
the action-at-a-distance theory for gravitational attraction, the
general consensus was broken again. It resulted in polarizing
views on both sides of the Channel: paradoxically, the majority of
English scholars were now on the side of Faraday's field
conception, whereas the notion of the direct action-at-a-distance
still remained fundamental on the Continent. \textit{Thus, the
19th century electrodynamics became the battle-field of a
paramount importance to test existing conceptions of interaction}.
There is no doubt that Hertz's contribution was decisive in that
historical choice. He was the first physicist who observed
electromagnetic waves in air travelling with the velocity of
light. Since then it is commonly believed that on the classical
level the question of the general meaning and understanding of
electromagnetic interactions and their mechanisms had been
positively resolved.

In this work we will discuss that Hertz's experimental approach on
propagation of electromagnetic interactions cannot be considered
as conclusive at many points as it is generally implied and,
hence, the central battle of the 19th century physics cannot been
considered definitely won by one of the opposing sides. This
conclusion depends essentially upon the correct reconstruction of
the historical retrospective of the 19th century's electromagnetic
research.

\section{Historical Retrospective of the 19TH Century's
Electromagnetism and its Conceptual Background}

In order to appreciate the difficulty and the importance of the
task undertaken by Hertz in his experimental investigations, it is
worth recalling the uncertain and highly controversial state of
electrodynamics of the 19th century. Hertz himself was trained in
the research tradition of the Berlin school headed by Helmholtz
$^{(6)-(7)}$, who from the middle of 1860's, had sought to clarify
existing principles in electromagnetic theory and to reach a
consensus between the two major directions in electromagnetic
research of that time, namely, Newton's instantaneous
action-at-a-distance concept as used by Weber, and Faraday's
contact action concept as developed by Maxwell. By the time of
Helmholtz's first attempt at reconciliation (1870), the
theoretical schemes of Weber and Maxwell had successfully
incorporated all previously well-established descriptions and
empirical facts, such as the electric potential theory
(electrostatics), Amp\`{e}re's magnetostatics and Faraday's theory
of induction.

Weber developed his theory (1848) in accordance with the Newtonian
program, which prescribed that all forces between pairs of
particles should be radial, acting directly through space (i.e.
along the line joining particles) without any observable material
mediator. Restriction on this radial description of
electromagnetic forces came from Amp\`{e}re, who understood that
instantaneity meant no delay, hence no aberration. Any aberration
attending the finite propagational velocity of interactions would
imply non-radial forces. However, the existence of radial forces
as a basic assumption of instantaneous action-at-a-distance
(IAAAD) theories was confirmed by Amp\`{e}re experimentally to the
degree of accuracy available at that time. Thus, electric and
magnetic interactions were thought to be complete analogy to
gravitational attractions which, according to astronomical
observations, always acted along the line joining simultaneous
positions of two bodies.

At first time, Faraday's material field concept attracted many
scholars seeking rather for the physical explanation than for the
purely mathematical description of gravitational and
electromagnetic phenomena. But later on, the perspective opened by
Faraday seemed to be so wide and novel that it challenged to
explore it deeply many excellent mathematicians of that time and
among them Maxwell, who also tried to work out his own
comprehensive field theory based on Faraday's concept. However,
Maxwell himself was aware of the provisional status of this
approximation and encountered some conceptual difficulties, since
he had incorporated all the basic IAAAD results such as
electrostatics and magnetostatics without any modification. With
regard to the lines of force treated by Faraday as the
representation of a material field, Maxwell's own position was
still undefined, but he cautiously dealt with them as if they were
lines of flow of an incompressible, imaginary fluid. As Maxwell
stated$^{(8)}$:

\begin{quotation}
\textit{The substance here treated of must not be assumed to
possess any of the properties of ordinary fluids except those of
freedom of motion and resistance to compression. It is not even a
hypothetical fluid which is introduced to explain actual
phenomena. It is merely a collection of imaginary properties which
may be employed for establishing certain theorems in pure
mathematics in a way more intelligible to many minds and more
applicable to physical problems than that in which algebraic
symbols alone are used.}
\end{quotation}

Consequently, being in a static limit, mathematically equivalent
to older IAAAD theories, the status of contact field theories
could not have been considered free of ambiguity (a fuller
discussion of this can be found in M. Hesse work$^{(9)}$). In
other words, in this static limit, electric and magnetic fields
behaved very much as flows of an ideal, incompressible fluid, in
which case they were indistinguishable from IAAAD. These
uncertainties in Maxwell's original theoretical scheme were later
summarized clearly and concisely by Hertz, who wrote$^{(10)}$:

\begin{quotation}
\textit{Maxwell's own representation does not indicate the highest
attainable goal; it frequently wavers between the conceptions
which Maxwell found in existence, and those at which he arrived.
Maxwell starts with the assumption of direct action-at-a-distance;
he investigates the laws according to which hypothetical
polarizations of the dielectric ether vary under the influence of
such distance-forces; and he ends by asserting that these
polarizations do really vary in this way, but without being
actually caused to do so by distance-forces. This procedure leaves
behind it the unsatisfactory feeling that there must be something
wrong about either the final result or the way which led to it.}
\end{quotation}

As an approach to clarifying these uncertainties, let us examine
the essential distinction between the conceptual foundations of
IAAAD and those of contact-field doctrines. At the beginning of
the 19th century, the possibility of a final explanation of the
mechanisms of IAAAD as the basis of gravitational and electric
forces had not been completely ruled out. Moreover, due to the
rapid development of theoretical hydrodynamics, the attitude
towards IAAAD changed from the summary rejection of its unphysical
status to an awareness of a deep similarity between the potential
function and the velocity-field of a fluid. It had been realized
that the main difference between IAAAD and Faraday's
\textit{field} was the fact that in IAAAD a potential need not
necessarily describe a material property of anything, whereas for
Faraday it was the property of a material substance which could be
observed in ways familiar to ordinary matter such as, for
instance, liquids and gases. Like any other material substance,
therefore, Faraday's $field$ could be regarded as something
movable with positive kinetic energy and, therefore, detectable
empirically. This suggests that an important criterion, as Hesse
puts it$^{(11)}$:

\begin{quotation}
\textit{... in deciding whether or not a field is to be regarded
as a physically continuous medium rather than a mere mathematical
device lies in its possession of detectable properties other than
the one property for which it was introduced. A condition of this
kind is often suggested as a of the physical 'reality' of a
theoretical entity, and it led Faraday to express his
dissatisfaction with Newtonian gravitation. But independent
detection was not the only consideration which weighed with the
nineteenth-century physicists. They were prepared to regard a
field as a physically continuous medium on other and less
stringent terms, for example, if propagation was affected by
material changes in the intervening space, if it took time, if a
mechanical model could be imagined for the action of a medium in
producing the observed effect, or if energy could be located in
the space between interacting bodies. Any of these three
conditions might be regarded as sufficient and no one of them was
individually necessary. Thus, gravitation remained an action at a
distance throughout the nineteenth century, in spite of its
description by a potential theory, because it did not satisfy any
of these criteria, whereas the electromagnetic field theory began
to take on the characteristics of continuous action because it
satisfied all of them. It is sometimes suggested in modern
(post-relativity) works that a finite velocity of propagation is a
necessary as well as sufficient condition for continuous action,
and that this is why instantaneous gravitation could not be
regarded as such an action, but in classical physics there is
instantaneous transmission of pressure and of longitudinal waves
in an incompressible medium, and this would certainly be regarded
as continuous action.} \end{quotation}

From this preliminary survey of the two opposite conceptual
foundations, we may conclude that the 19th century physicists
considered as sufficient conditions for validity of
Faraday-Maxwell's field concept:

1) \textit{propagation of fields producing material changes in the
surrounding space};

2) \textit{time-delay in propagation};

3) \textit{experimental observation of the energy transfer related
to interactions};

Nevertheless, as it will be discussed later, the above-mentioned
conditions should be considered only as necessary, not sufficient
for establishing the existence of Faraday-type contact fields,
since there might be a third alternative which would combine both
IAAAD and Faraday's contact-field features in a single scheme.
Formally, such a scheme (with superposition of \textit{material}
and \textit{immaterial} substances) would satisfy all above
mentioned conditions but on theoretical foundations different from
those of purely \textit{material} contact-field theory. In this
respect, it will be shown when examining the historical background
of the 19th century electrodynamics, that such a third alternative
did actually exist and that this was the so-called compromise
theory of Helmholtz. It means that the problem of the completeness
of Hertz's $crucial$ experiments should be again revisited but
from methodological and philosophical positions of modern science.

\section{Status of Different Electrodynamics Doctrines Before
Hertz's Experiments}

By the time Helmholtz became actively involved in resolving
problems of electromagnetism, in the middle of the 1860's, Weber's
and Maxwell's supporters had already been locked in a lengthy and
futile conflict. Helmholtz attempted to make a decisive choice
between them by constructing his own mathematical scheme and
designing crucial experiments to weigh in favor of either Weber's
or Maxwell's theory.

Helmholtz$^{(12)}$ attempted to elaborate his compromise approach
aimed at combining the important elements of the two theories.
When trying to arrive at results similar to Maxwell's without
losing the elements of action-at-a-distance, Helmholtz assumed
that the electrostatic forces are constantly present as a field in
space and that the change in the polarization or the displacement
of the charges signalled the change in the electrostatic field. As
discussed by Kudryavtzev$^{(13)}$ and by Buchwald$^{(14)}$, under
these assumptions, Helmholtz successfully derived generalized
equations very similar to those of Maxwell and found that in a
limit case they yielded equations identical to Maxwell's equations
for scalar and vector potentials. Solving these equations for a
homogeneous dielectric medium, he arrived at the wave equations
for electric and magnetic polarizations, respectively, with
undetermined constant \textit{k} (see$^{(13)-(15)}$).

The conciliatory aspect of Helmholtz's approach resulted in the
following peculiarity. To reach formally Maxwell's scheme from
Helmholtz's approach required two limits: \textit{k=0} and
infinite ether susceptibility (for greater details see Appendix
A). The former was necessary to obtain Maxwell's equations for
potentials and the latter, to make the velocity of transverse
waves in Helmholtz theory equal to the velocity of light
\textit{c}. In addition to the ordinary transverse electromagnetic
waves already confirmed by Maxwell, Helmholtz discovered the
existence of longitudinal electric waves which turned out to be
instantaneous at the Maxwellian limit. Interpretation of this
conclusion and its consequences became a hard nut to crack for all
contemporary electrodynamicists. Maxwellian followers (Heaviside,
FitzGerald, Lodge etc.) refused to accept Helmholtz's theory
because they found his conceptions entirely foreign to Maxwell's
view of the transmission of interaction. Helmholtz himself
recognized conceptual differences between the two theories but
insisted on their remarkable similarity$^{(16)}$:

\begin{quotation}
\textit{...It follows. ...from these investigations that the
remarkable analogy between the motion of electricity in a
dielectric and that of the light ether does not depend on the
particular form of Maxwell's equations hypotheses, but results
also in a basically similar fashion if we maintain the older
viewpoint about electrical action at a distance.}
\end{quotation}

Helmholtz's attempt at a more consistent reformulation of the
contemporary electrodynamics theories could not, however, resolve
the problem of which approach to favor. His theory contained the
undefined constant \textit{k} which could have any possible
non-negative value and still remain compatible with existent
experimental data, since most of it had been obtained for closed
currents where the value of \textit{k} is irrelevant. Even for
values of \textit{k} close to unity in the case of open currents,
an experimental effort towards obtaining reliable results would
require an enormous boost in accuracy and would therefore require
new kinds of measurement devices. Despite the difficulties in
carrying out any experimental task of this kind, Helmholtz, along
with N. Schiller and H. Rowland, his pupils and followers at the
University of Berlin, performed in 1874-78 a series of preliminary
experiments on open currents. All results were in apparent
agreement with the law of electromagnetic induction described
equally well by the Neumann, Maxwell and Helmholtz theories.
Therefore, the need for new decisive and reliable experimental
data was still urgent when Hertz became interested in
electromagnetic research.

\section{Hertz's Contribution to Electromagnetic Research}

In 1879 Helmholtz proposed a prize competition, "\textit{To
establish experimentally a relation between electromagnetic action
and the polarization of dielectrics}" and urged his student Hertz
to take up the challenge. At first, Hertz declined, discouraged by
the poor prospects of success at that time. Later on, he began
investigating the problem for his own interest. In 1886-88, at
Karlsruhe, he attempted to establish the compatibility of the
theories of Helmholtz and Maxwell in a new series of experiments.
He designed his measurement-procedure, taking into account
Helmholtz's separation of the total electric force into the
electrostatic and electrodynamic parts to which different
velocities of propagation were ascribed. In Hertz's
words$^{(17)}$:

\begin{quotation}
\textit{The total force may be split up into the electrostatic
part and electrodynamic part; there is no doubt that at short
distances the former, at great distances the latter, preponderates
and settles the direction of the total force.}
\end{quotation}

In spite of apparent difficulties, Hertz decided to carry out
those measurements. For his purpose he had to develop new devices
providing electrical oscillations faster than had been previously
available. His efforts were, however, rewarded and he quickly
demonstrated the existence of very rapidly varying currents with a
strong inductive action across the discharge gap$^{(18)}$. He also
established a resonance-like relationship between the primary and
secondary circuits in the presence of regular oscillations. As a
result of these investigations, a solution to the Berlin Academy's
problem could be considered possible.

The next important step was to modify and improve the detection
apparatus (secondary circuit) in order to undertake interference
experiments in which he used mutually orthogonal wire and
oscillator. Hertz's idea was the following: the oscillator sends
out a given phase along the wire as well as a given phase of
direct action through air (which differs from the wire phase by a
quarter of a period). If both phases travel at the same rate,
their combined effect (interference) will be the same along the
wire. Otherwise, the effect will change at different point.

At the end of 1887, On November 11, he finally undertook this task
and observed the existence of the interference between the
oscillations propagated in air and wire. The entry of his dairy
notes for November 12 reads: "\textit{Set up experiments on the
velocity of propagation of the electromagnetic effect. Contrary to
expectations, the result is infinite propagation}". In
retrospective, Hertz gave account about these first experiments in
the Introduction to his "Electric Waves":

\begin{quotation}
\textit{...Dishearted, I gave up experimenting. Some weeks passed
before I began again. I reflected that it would be quite as
important to find out that electric force was propagated with an
infinite velocity, and that Maxwell's theory was false, as it
would be, on the other hand, to prove that this theory was
correct, provided only that the result arrived at should be
definite and certain...}
\end{quotation}

As we know from his dairy, he decided to write a detailed paper on
negative results but only after having checked and ensuring his
previous results. Five weeks later, On December 15-22, Hertz
returned to his experiments and confirmed apparently infinite
propagation of direct action (On the sequence of Hertz's first
experiments, see Appendix B).

On December 26 Hertz began a new series of experiments using a new
position of the resonator. According to the measured interference
phase shift, Hertz unexpectedly found that the oscillator's direct
action propagated $1.5$ times faster than the wire wave (i.e.
close to the velocity of light). Setting the resonator in
different positions Hertz obtained interference marks which he put
in the final table. As we shall discuss below, these results
contained a singularity: an apparently infinite rate of
propagation in region very close to oscillator (near zone).

After some doubts (see Appendix B) Hertz compiled all result of
the last experiments and published in a paper entitled "\textit{On
the Finite Velocity of Propagation of Electromagnetic Action"
(1888)} which nowadays, according to the established
historiography of physics, is considered a classical reference in
which the difficult task of proving the finite propagation
velocity of electromagnetic interactions in air had been achieved
(it should be noted here that the title of this Hertz's paper is
perhaps misleading nowadays, because conventional Maxwellian
electrodynamics does not employ the Helmholtzian \textit{action}
terminology, nor does it split the total electric force into
electromagnetic and electrostatic parts).

It is important to understand what Hertz had in mind when
published this paper. As we know from Hertz's dairy and letters,
he was eagerly looking for a positive result, i.e completely novel
effects (see Appendix B). After the last series of experiments he
became strongly convinced the electrodynamic action had a finite
rate of propagation. However, he was not sure how to combine it
with the previous experiments and found it possible to question
their viability in the new light of novel "\textit{positive
results}". In fact, he admitted that his first experiments could
not be considered reliable to conclude something definitive about
longitudinal components. Looking carefully through the same paper,
we find Hertz declaring to support his claims$^{(19)}$:

\begin{quotation}
\textit{From this it follows that the absolute value of the first
of these is of the same order as the velocity of light. Nothing
can as yet be decided as to the propagation of electrostatic
actions.}
\end{quotation}

In later retrospective, writing the Introduction to his "Electric
Waves" Hertz argued apparent implications of his "negative"
results by inaccuracy in observations (for more details see
Hertz's Introduction to "Electric Waves and also Appendix B).

Concerning the final table of interference marks which he
published in the paper "\textit{On the Finite Velocity of
Propagation of Electromagnetic Action" (1888)}, it should be
emphasized that some of Hertz's measurements tended to manifest
the instantaneous nature of the electrostatic mode, but he was
also not convinced of this instantaneity and preferred to be
cautious, since his method was unable to provide him with any
reliable quantitative results$^{(20)}$:

\begin{quotation}
\textit{Since the interferences undoubtedly change sign after 2.8
meters in the neighborhood of the primary oscillation, we might
conclude that the electrostatic force which here predominates is
propagated with infinite velocity. But this conclusion would in
the main depend upon a single change of phase... If the absolute
velocity of the electrostatic force remains for the present
unknown, there may yet be adduced definite reasons for believing
that the electrostatic and electromagnetic forces possess
different velocities.}
\end{quotation}

These circumstances of Hertz's experiments were previously noted
by some researchers on history of physics such as
Cazenobe$^{(21)}$ and Doncel$^{(22)}$. An excellent reconstruction
of Hertz's "crucial" experiments in great detail was undertaken by
Buchwald$^{(14)}$ where one can appreciate some of Hertz's major
doubts and hesitations (see Appendix B).

In spite of all difficulties and uncertainties of these first
measurements, Hertz was fully aware of the need for additional
experiments to cast some light of certainty on the electrostatic
part$^{(19)}$:

\begin{quotation}
\textit{It is certainly remarkable that the proof of a finite rate
of propagation should have been first brought forward in the case
of a force which diminishes in inverse proportion to the distance
[electrodynamic part], and not to the square of the distance
[electrostatic part]. But it is worth while pointing out that this
proof must also affect such forces as are inversely proportional
to the square of the distance. For we know that the ponderomotive
attraction between currents and their magnetic actions are
connected by the principle of the conservation of energy with
their inductive actions in the strictest way, the relation being
apparently that of action and reaction. If this relation is not
merely a deceptive semblance, it is not easy to understand how the
one action can be propagated with a finite and the other with an
infinite velocity.}
\end{quotation}

Hertz's point seems to tell his growing insatisfaction with the
complexity of Helmholtz's theoretical approach. In his opinion,
two different velocities ascribed in Helmholtz's theory to two
different parts of action made the whole task of test and analysis
unreasonably complicated. Hertz inclined towards justification of
Faraday-Maxwell's field approach as a special limit of Helmholtz's
theoretical scheme and also as based on a simpler $model$.

Hertz's conversion to Maxwellian ideas is well-discussed in some
history and philosophy of science texts$^{(23)}$. Nevertheless, it
is interesting to follow briefly the evolution of Hertz's
theoretical ideas. He started with Helmholtz's theory, and his
conversion to Maxwell's viewpoint was an uneasy process, possibly
never fully completed due to his (Hertz's) premature death (On
January 1, 1894). He began analyzing the underlying concepts in
the Maxwellian limit of Helmholtz's theory (see Appendix A) but
his final interpretation became essentially different in form from
what had been commonly accepted by Helmholtz and his followers.
More specifically, Hertz uncritically assumed that in Maxwell's
limit the instantaneous longitudinal component should have been
excluded from consideration in Helmholtz's original theory. All
forces then became explicitly time dependent, i.e. possessing the
finite velocity of propagation.

This was a drastic departure from the Hertz mentor's position on
the nature of electromagnetic interactions and, in general, from
his philosophical foundations. In fact, Helmholtz rejected time
dependent forces (he admitted only implicit time dependence upon
space position or direct action-at-a-distance (IAAAD)) and was
deeply convinced that$^{(24)}$
\begin{quotation}
\textit{...nature could only be comprehended through invariable
causes. Helmholtz viewed electromagnetic interactions - indeed,
all interactions, - as instantaneous and bipartite...}
\end{quotation}
and, therefore, could attribute interactions only to longitudinal
components (electrostatic action in Helmholtz's classification).

Let us see an aberration of Helmholtz's ideas in Hertz's own
words$^{(25)}$:
\begin{quotation}
\textit{...Helmholtz distinguishes between two forms of electrical
force the electromagnetic and the electrostatic to which, until
the contrary is proved by experience, two different velocities are
attributed. An interpretation of the experiments from this point
of view could certainly not be incorrect, but it might perhaps be
unnecessarily complicated. In a special limiting case Helmholtz's
theory becomes considerably simplified, and its equations in this
case become the same as those of Maxwell's theory; only one form
of the force remains, and this is propagated with the velocity of
light. I had to try whether the experiments would not agree with
these much simpler assumptions of Maxwell's theory. The attempt
was successful. The result of the calculation are given in the
paper on "The Forces of Electric Oscillations, treated according
to Maxwell's Theory".}
\end{quotation}

This paper was published in 1889, one year after the discussion of
Hertz's first results, which apparently were not sufficient to
conclude which of the two theoretical descriptions was more
adequate. Hertz evaded the wide analysis of the nature and
mechanisms of electromagnetic interactions, reducing it to the
limited task of experimental verification of some predictions of
Maxwell's theory. In this comprehensive paper Hertz tried to show
how the observed singularities in the propagation of the electric
force could be described by Maxwell's theoretical scheme. As Hertz
explained$^{(26)}$:
\begin{quotation}
\textit{The results of the experiments on rapid electric
oscillations which I have carried out appear to me to confer upon
Maxwell's theory a position of superiority to all others.
Nevertheless, I based my first interpretation of these experiments
upon the older views, seeking partly to explain phenomena as
resulting from co-operation of electrostatic and electromagnetic
forces. To Maxwell's theory in its pure development such a
distinction is foreign. Hence I now wish to show that the
phenomena can be explained in terms of Maxwell's theory without
introducing this distinction. Should this attempt succeed, it will
at the same time settle any question as to a separate propagation
of electrostatic force, which is meaningless in Maxwell's theory.}
\end{quotation}

In this famous paper, Hertz wrote Maxwell's equations in the form
in which they are known today (the Hertz-Heaviside form) and also
derived the distribution of force lines for the radiating
oscillator (Hertz vibrator). In other words, this important
contribution to the Faraday-Maxwell field theory consisted in the
development of the general source-field relation previously
unknown. (Today this method bears Hertz's name and is based on the
straightforward Fourier analysis of dipole and multi-dipole
radiation.) Using these calculations, Hertz found an explanation
(alternative to that based exclusively on Helmholtz's ideas) of
the singularities he had observed in the distribution of radiation
in the near field (the apparently instantaneous behavior of the
electrostatic component). In Hertz's words$^{(27)}$:
\begin{quotation}
\textit{Let us now investigate whether the present [Maxwell's]
theory leads to any explanation of the phenomena... At great
distances the phase is smaller by the value $\pi$ than it would
have been if the waves had proceeded with constant velocity from
the origin; the waves, therefore, behave at great distances as if
they had travelled through the first half wavelength with infinite
velocity.}
\end{quotation}

Interestingly, this prediction of Maxwell's theory concerning the
infinite phase-velocity for the near-field zone based on
straightforward Fourier analysis (Hertz's method) appears in
modern texts$^{(28)}$. Surprisingly, however, that Hertz himself
paid no attention to this prediction beyond the fact that it gave
him a new interpretation of his experimental results, different
from that provided previously only by the Helmholtzian approach.
It is possible that Hertz did not realize (or had no time to
realize) all conceptual implications of the new prediction, which
has no clear meaning in the framework of the Faraday-Maxwell
contact-field doctrine. The prediction would imply the existence
of a small, but macroscopic region where the notion of Faraday
locality results invalid. On the other hand, it reproduces
above-discussed \textit{fuzziness} in the relationship between
static and dynamic limits in Maxwell's theory.

Bearing that ambiguity in mind, it would not have been surprising
if Maxwell's theoretical predictions for static and quasistatic
phenomena had been found to be similar to the older IAAAD views.
It is obvious that phenomena in the near field zone (less than
half wavelength) should be regarded as quasi-static in the
Hertzian analysis and therefore, implicit time dependent that is
not Maxwellian but Helmholtzian feature for longitudinal
components. This reasoning should have cast doubts on Hertz's
explanation of the experimental results as not being completely in
the spirit of the Faraday-Maxwell's conceptual foundations. It is
surprising, then, that almost no-one seemed to have been worried
by this presence of non-Faraday's elements in Maxwell's approach.
By the same token, it is no less surprising that Hertz's
explanation was so unconditionally accepted by Maxwell's
followers.

Although Hertz was satisfied that his calculations had accounted
for the majority of the observed phenomena, he stressed that he
had not succeeded in removing all the difficulties from his
experimental verification of Maxwell's theory. He confessed
that$^{(29)}$:
\begin{quotation}
\textit{I have therefore repeated the experiments, making various
alterations in the position of the primary oscillator, and found
that in certain positions the results were in accordance with
theory. Nevertheless, the results were not free from ambiguity,
for at great distances in places where the force was feeble, the
disturbance due to the environment of the space at my disposal
were so considerable that I could not arrive at a trustworthy
decision.}
\end{quotation}

This shows that some of Hertz's measurements had been made to the
limit of accuracy available at the time. In his final experimental
paper titled "\textit{On Mechanical Action of Electric Waves in
Wires}", published in 1891 (after his departure from Karlsruhe),
Hertz attempted to observe the existence of magnetic waves
accompanying the electric waves in order to disprove any reference
to action-at-a-distance. As far as the problem of electromagnetic
interaction is concerned, this last task was an important attempt
on a definite experimental demarcation of irreconcilable views.
However, he was forced to admit$^{(30)}$:
\begin{quotation}
\textit{...I hoped to be able to devise some way of making
observations on waves in free air, that is to say, in such a
manner that any disturbances which might be observed could in no
wise be referred to any action-at-a-distance. This last hope was
frustrated by the feebleness of the effects produced under the
circumstances}.
\end{quotation}

Additionally, there were no empirical indications of the energy
transfer between interacting electrically charged bodies at rest.
The question remained open whether Maxwell's transverse waves of
interaction were too feeble to be detected or, otherwise, they had
no right to exist according to Helmholtz's views. Thus, Hertz
himself recognized that the fundamental question on the nature and
mechanisms of interaction could not have been definitively
clarified by experimental tools available at that time.

Although Hertz's satisfaction with Maxwell's theory was
understandable, there were still insufficient number of arguments
for making a truly decisive choice, bearing in mind that
Helmholtz's approach remained in qualitative agreement with the
observed singularities. However, no further experiments or
calculations for testing the quantitative predictions of
Helmholtz's theory have been attempted.
\begin{quotation}
\textit{...The problem Hertz encountered with Helmholtz's theory
was thus of a very special sort. It does not consist in a
contradiction between an empirical observation and theory, nor
does it consist in a contradiction among theoretical statements.
Instead it consists in a philosophical problem with theoretical
terms: what is their meaning, how do they acquire their meaning,
how can they be rendered consistent}.$^{(31)}$
\end{quotation}

In sum, we can conclude that no empirical indication of the
inadequacy of Helmholtz's theory had ever been observed or
declared by Hertz, so that from the position of modern scientific
method it is difficult to conceive why Helmholtz's theory had been
ruled out as a possible alternative. In a mathematical language,
we would say that Hertz found only necessary conditions to accept
Maxwell's theory. If those conditions were sufficient as well,
then Helmholtz's theory would have been discarded on a solid
logical basis. In the next section, necessary and sufficient
conditions for a decisive choice on the adequate account of
electromagnetic phenomena and interactions will be discussed.

\section{Discussions}

In modern retrospective of the above-stated reservations of Hertz,
as well as of some other thinkers of the time, the unconditional
acceptance with which Hertz's experiments and their
interpretations were received might seem somewhat unjustified.
Hertz himself did not expect such support and attributed it to the
heavy philosophical burden of the old and unresolved dilemma of
the choice between the IAAAD and contact-action
doctrines$^{(32)}$:
\begin{quotation}
\textit{The approval with which they have been received has far
exceeded my expectations. A considerable part of this approval was
due to reasons of a philosophic nature. The old question as to the
possibility and nature of forces acting at a distance was again
raised. The preponderance of such forces in theory has long been
sanctioned by ordinary common sense; in the domain of electricity
these forces now appeared to be dethroned from their position by
simple and striking experiments... The details of the experiments
further prove that particular manner in which the electric force
is propagated exhibits the closest analogy with the propagation of
light; indeed, that it corresponds almost completely to it.}
\end{quotation}

There is no doubt, then, that Hertz already considered the
propagation of electric interaction as a transverse wave,
completely analogous to the light-propagation, due to its being
the only explicit indication of his experiments. Another detail
which may also have contributed to the unconditional approval of
Hertz's results among the scientific community was, as already
stated, the rather misleading title of his paper on the finite
propagation of electromagnetic actions, possibly due to
unawareness of Helmholtz's classification of forces.

However, in contrast to this general enthusiasm for Hertz's
results, there was also some strong opposition from P. Duhem, an
eminent French mathematician, physicist and philosopher of science
at the beginning of the 20th century. He was one of a small but
honest group of scientists who refused to accept Hertz's
experiments as conclusive. Moreover, Duhem was the first to raise
doubts about the whole concept of $crucial$ experiments$^{(33)}$.
A good mathematician and outspoken critic of the inconsistencies
in Maxwell's theory, he became one of the principle advocates of
Helmholtz's approach$^{(34)}$:
\begin{quotation}
\textit{...Physicists are caught in this dilemma: Abandon the
traditional theory of electric and magnetic distribution, or else
give up the electromagnetic theory of light. Can they not adopt a
third solution? Can they not imagine a doctrine in which there
would be a logical reconciliation of the old electrostatics, of
the old magnetism, and of the new doctrine that electric actions
are propagated in dielectrics? This doctrine exists; it is one of
the finest achievements of Helmholtz; the natural prolongation of
the doctrines of Poisson, Amp\'{e}re, Weber and Neumann, it
logically leads from the principles laid down at the beginning of
the nineteenth century to the most fascinating consequences of
Maxwell's theories, from the laws of Coulomb to the
electromagnetic theory of light; without losing any of the recent
conquests of electrical science, it re-establishes the continuity
of tradition.}
\end{quotation}

However, it appears that this call of Duhem's for a \textit{third
solution} fell mainly on deaf ears. So, whilst appreciating the
difficulties of Hertz's pioneering investigations, and taking into
consideration his struggle through the uncertainties and
controversies of the electrodynamics of his time, the fact remains
that his final opting for Maxwell's theory was not based on strict
scientific logic. What needs to be done, therefore, is clearly to
identify the criteria for acceptance and apply these to the
existing alternative theories. The detailed examination of how and
why a certain theory is confirmed or refuted by experimental tests
is, of course, a matter of the methodology and philosophy of
science. Hertz was apparently not fully aware of the need to test
his choice from this systematic, methodological and philosophical
standpoint.

As did the majority of his contemporaries, Hertz intuitively
applied a criterion of empirical verification, in the
hypothetico-deductive manner prevailing in the 19th century
science. This method consisted of creating hypotheses in the form
of postulates and then making deductions from these which could be
either confirmed or rejected by experiment. However, from the
modern methodological standpoint, as is now well recognized,
empirical verification is a condition that is necessary but not
sufficient for establishing the truth of a theory. The fact is
that an empirical observation may $verify$ any number of
different, yet equally valid theories sharing that same
prediction. The only reliable way, therefore, of deciding between
theories is not to $verify$ any one of them in particular, which
might just as well verify any number of them but, if possible, to
eliminate all but one of the contenders. The only way of doing
this is, of course, by the method of refutation. This method may
be carried out logically, by pure ratiocination (as by pointing
out some logical or mathematical contradiction) or by
demonstrating that empirical predictions made by some logically
consistent theory are false. However, even if a theory is
logically sound, if it makes no falsifiable predictions, it cannot
be refuted. Such a theory, according to Popper, cannot qualify as
a scientific theory. This criterion, introduced by Popper in the
late 1920s, thus provided a reliable criterion for separating
genuine scientific theories from pseudo-scientific or
$metaphysical$ theories by which, Popper meant theories which make
no predictions that can, even in principle, be empirically
falsified.

It may be of interest to consider how this modern criterion of
falsifiability might have been influenced on Hertz's decision, had
it been available at the time. Qualitatively, as we have seen,
both Maxwell's and Helmholtz's theories fit equally well all the
observations made by Hertz, namely:

1) \textit{material changes in the surrounding space};

2) \textit{finite propagation of transverse components with the
velocity of light};

3) \textit{empirical observations of energy transfer etc}.

The present-day scientific method suggests that the next step is
to explore the difference between the experimental predictions of
the two theories, beyond those that are already known, and to
separate-out the different, non-compatible but empirically
verifiable predictions that is, to determine which of them, if
any, are sufficient to explain all known facts, as distinct from
those that are merely necessary. Thus, in the case of alternative
electrodynamics theories, the core of the additional
'\textit{crucial}' experiment should have been to test,
experimentally, statements specifically describing the character
of the longitudinal electric components which distinguish the
Helmholtzian from the Maxwellian theory. A clear absence of any
indications on longitudinal character of interactions in all
observable cases would immediately imply Maxwell's theoretical
approach beyond any doubt. Since no decisive, unambiguous
information of the kind necessary to refute Helmholtz's theory has
yet been found in Hertz's experimental results, and, hence no
sufficient criterion had been established to accept the
Faraday-Maxwell field interpretation. Moreover, as we mentioned
above, electrostatic interactions always act along the line
joining positions of two charges at rest, i.e. they are manifestly
longitudinal. Proper Maxwell's equations give the same picture in
electro- and magnetostatic limits. The point is that
longitudinality of interactions is an essential, irremovable
feature in many observable phenomena. There was undoubtedly a real
difficulty here, realized by Hertz himself after frustrated
efforts to dissociate any reference to IAAAD as far as to
interaction was concerned.

The criterion on empirical observations of energy transfer comes
from Faraday. He believed that any interaction should be
accompanied by material changes in the surrounding space, that its
propagations should take time. According to this reasoning, the
alternative views would violate the conservation of energy because
it would appear in the body without passing the intervening space.
Later on Poynting developed this conception in deeper details, so
that the energy could be seen in many respects similar to a
material fluid satisfying equations of conservation and
continuity. The flux of interaction was since then up till now
associated with the flux of electromagnetic field energy.
Nevertheless, it should be recognized that there are still various
well-known difficulties in modern electrodynamics in describing
the energy flow in some physical situation.

On the contrary, Helmholtz's position differentiated both kinds of
energy: potential and kinetic as it was done in mechanics. The
energy of transverse waves represented material kinetic energy
localized in space between radiating bodies whereas potential
energy had no material equivalent and did not show any property of
its manifest accumulation in the intervening space (non-local and
non-material property). Thus, in Helmholtz's interpretation
longitudinal components (responsible for interaction) did not take
part in any energy carrying through the space. \textit{Immaterial}
status of longitudinal instantaneous forces enable them to avoid a
possible conflict with the special relativity. In fact, it is
well-known that Einstein's theory does not limit phase velocities,
if there is no local energy transfer.

Before long physicists ceased to ask why no scientific device is
able to detect any observable energy flow between two interacting
bodies at rest. Recent photon teleportation experiments also
questioned actual understanding of localization of electromagnetic
energy. These difficulties seem to arise in modern physics from
the indispensability of material nature of all kinds of
electromagnetic and gravitational interactions. This comes from
Faraday and Maxwell who thought that if the IAAAD concept did not
imply any material mediator, then it did not imply any conceivable
physical mediator at all. Therefore, in respect to the energy,
they thought it could have only material origin, being, like
matter, accumulated locally and transferred continuously in space.
This path of reasoning also explains the modern prejudice against
immateriality, even though space and time can hardly be classified
among material objects and hence cannot possess only $material$
properties. Consideration of both kinds of electromagnetic energy
following Helmholtz (local along with non-local, i.e. material
kinetic along with potential) helps to overcome this fundamental
methodological inconsistency but on theoretical foundations
different from those of purely $material$ Faraday-Maxwell's field
theory.

On the other hand, as already mentioned, the criterion of
falsifiability requires any truly scientific alternative theory to
be logically self-consistent. Logical inconsistencies lead to
bogus predictions. In this way it is interesting to remind that
several aspects of standard conventional electrodynamics are found
to be unsatisfactory, despite all the advances claimed by
relativity and quantum mechanics. Conventional electrodynamics is
thus still not free from untractable inconsistencies, as in its
implications regarding self-interaction, infinite contribution of
self-energy, the concept of electromagnetic mass, indefiniteness
in the flux of electromagnetic energy, etc. These internal
difficulties explain why, from the beginning to the middle of the
20th century, there were unceasing efforts to modify either
Maxwell's equations or the underlying conceptual premises of
electromagnetism. The present status of classical electrodynamics
can be expressed by words of R. Feynman$^{(35)}$:
\begin{quotation}
\textit{...this tremendous edifice [classical electrodynamics],
which is such a beautiful success in explaining so many phenomena,
ultimately falls on its face. When you follow any of our physics
too far, you find that it always gets into some kind of trouble.
the failure of the classical electromagnetic theory. ...Classical
mechanics is a mathematically consistent theory; it just doesn't
agree with experience. It is interesting though, that the
classical theory of electromagnetism is an unsatisfactory theory
all by itself. There are difficulties associated with the ideas of
Maxwell's theory which are not solved by and not directly
associated with quantum mechanics...}
\end{quotation}

One of the latest systematic accounts of these difficulties has
been made recently in$^{(36)-(37)}$. In particular, the old
problem of Maxwell's electrodynamics, concerning the uncertain
relationship between static and dynamic limits, has been brought
into greater relief. Pure mathematical analysis$^{(36)}$ has shown
that the conventional theory does not ensure a continuous
transition between static and dynamic limits, which is, surely, in
itself, strange to contemplate in the context of the
Faraday-Maxwell continuous-field concept. Interestingly, it has
also been found that if the condition of continuous transition
between static and dynamic limits is imposed explicitly in
mathematical terms, then conventional boundary condition should be
replaced by new generalized boundary conditions. As a result,
solutions of Maxwell's equations have to be also modified to
include instantaneous longitudinal components. It would be of
interest to note that the possible difficulty with conventional
boundary conditions for field equations in the classical field
theory was realized by A. Einstein himself a few month before his
death in 1955. In the last edition of "\textit{The Meaning of
Relativity}" he added the following$^{(38)}$:

\begin{quotation}
\textit{A field theory is not yet completely determined by the
system of field equations. Should one postulate boundary
conditions? Without such a postulate, the theory is much too
vague. In my opinion the answer to the question is that
postulation of boundary condition is indispensable.}
\end{quotation}

Thus, a different choice of boundary conditions can result in
essentially different approach (Maxwellian or Helmholtzian),
provided the same system of Maxwell's equations (a limit case of
Helmholtz's equations).

In summarizing this discussion it is interesting to note another
possible attractiveness of Helmholtz's conceptual foundations. As
well as having these above-mentioned difficulties, Maxwell's
theory also could not provide any reliable model of the atom. This
left a theoretical gap which had to be filled independently. This,
of course, is what led to the current quantum mechanical theory of
the atom, with its implications of essential non-locality. Thus
was created the other serious dilemma of present-day physics:
significant incommensurability between the classical relativistic
theory with its basic concept of local material field, and quantum
mechanics with its essential non-locality. This fundamental
conflict is such that, for some people, the only way of resolving
it seems to be to combine or superimpose these incompatible
requirements for locality and non-locality in some purely
expedient and compromising way. In view of these reasoning it also
can be suggested that any physical theory of interaction might not
be pure field theory (in the sense of local field) but be
complemented by instantaneous longitudinal components (in the
sense of non-local potential field).

\section{Conclusions}

Thus, as the relevant historical literature shows, Helmholtz's
theoretical foundations never contradicted Hertz's experimental
observations. Hertz's arguments in favor of Maxwell's theory had
not been based on a solid scientific and logical grounds as it has
been shown from the position of modern science and from the
analysis of very recently published Hertz's laboratory notes,
dairy and private communications$^{(22)}$.

Nowadays, Helmholtz's conceptual foundations promise an
alternative consistent solution to fundamental problems of modern
physics at reconciling classical electrodynamics and quantum
mechanics in a less ad hoc and altogether more rational way than
has, up till now, seemed obligatory. Recent experimental
confirmations of the violation of Bell's inequalities in quantum
mechanical measurements, entanglement and teleportation in quantum
optics shed some new light on possible alternative foundations of
classical electrodynamics.

The crisis which arose in classical physics at the very end of the
19th century accounted to the limitations of classical explanation
did not implied, however, radical revision of all foundations and
left unquestionable many previous results: among them conceptual
foundations of Maxwell's theory and Hertz's incomplete
experimental approach. Many problems of classical electromagnetic
research passed unnoticed in view of novel and promising
perspectives opened by special relativity and quantum mechanics.
Nevertheless, nowadays it is still unclear by what reasons the
progress of science at the beginning of the 20th century was so
unconditionally left to be dependent on formally incomplete
experimental verification of fundamental issues as far as to the
propagation of electromagnetic interactions is concerned.

Despite the intensive and undeniable progressive development of
physics and technology since Hertz's experiments, there is,
however, an uneasy feeling that we may not be on a very solid
ground with respect to the nature of interactions. In fact, modern
explanation of interaction process as an interchange of real and
virtual particles is still incomplete, posing more fundamental
questions than answers. In this respect, the standard model has
very high expectation on observations of fundamental Higgs's
massive particles in expensive CERN's experiments (approximately
in 2005). The failure would mean a possible crash of the
conventional approach. Why not to choose less expensive but not
less convincing way: using modern technological advantage over
Hertz's experimental facilities, it would be relatively easy to
complete Hertz's empirical task definitively and find out whether
the propagation of electromagnetic interactions is only transverse
(by means of massless photons), only longitudinal (Helmholtz's
non-locality) or both by nature on the classical level. This
latter task can also be considered as a sufficient condition for
verification of Maxwell's or Helmholtz's foundations.

\begin{center}
\textbf{APPENDIX A. Foundations and Structure of Helmholtz's
Theory}
\end{center}

The original Faraday-Maxwell form of electrodynamics which was
predominant in Britain, dispensed altogether with the objects like
charge in their own right. The primary entity turned out to be the
\textit{field}. The notion of a charge was very problematic in
Maxwell's original theory that made the electromagnetic fields to
be divorced from their sources. In Buchwald's words, "\textit{in
Maxwellian theory, charge is produced by the electric field...The
Maxwellian goal was to create a theory of electromagnetism which
made no use whatsoever of the microstructure of matter}" (J.
Buchwald, "\textit{From Maxwell to Microphysics}", The University
of Chicago Press, 1985). In other words, in the original field
theory the source was a secondary conception conceivable as a sort
of field singularity. In modern notation, this approach leads to a
particular form of continuity equation with no open currents
$\mathbf{J}$: (i.e. $\nabla \cdot \mathbf{J}=0$). (Later on the
notion of a source in Maxwell's theory was changed by Lorentz and
used in his microscopic electron theory of electromagnetism
accepted nowadays). The interaction between objects was reduced to
the change of state of the surrounding field. Hence, no direct
interaction was admitted in Faraday-Maxwell's electrodynamics as
in any other possible field theory, based on the notion of local
Faraday field. This marked the main division between Maxwell's and
Helmholtz's conceptions.

Helmholtz created a theory that differed radically from the
Faraday-Maxwell scheme. He did not see it appropriate to use
additional theoretical '\textit{artifact}', i.e. the Faraday
notion of local field. On the contrary, Helmholtz accepted
(following the main trend of the continental electromagnetism) the
underlying electric microstructure of matter as a primary
hypothesis and, hence, used a continuity equation that linked
charge and currents accepted in our modern notations. On the other
hand, he based his analytical approach on a notion of potential
functions. Besides the notion of a static potential (nowadays
known as a scalar potential function) satisfying inhomogeneous
Poisson's equation, Helmholtz postulated the existence of
electrodynamic potential $\mathbf{U}$ (nowadays known as a vector
potential function) that generalized the conception of a potential
so successfully used in magnetostatics. The source (charge or
current) specified in Helmholtz's approach the nature of
interaction (electric or magnetic). All objects interacted
directly, all forces among them could be deduced in terms of
potential functions and hence there was no intermediary local
field to interact with. Another essential point to grasp about the
fundamental structure of Helmholtz's electrodynamics is that every
system state can be described by the energy of interaction based
on both electrodynamic and static potentials. It reflects
Helmholtz's conviction in the profound truth of the
\textit{Principle of Least Action}. (Modern field theory is also
coupled to Hamilton's principle).

Helmholtz considered the total current $\mathbf{C}_{tot}=\mathbf{C}_{cond}+%
\frac{\partial \mathbf{P}}{\partial t}$ to be formed of conduction currents $%
\mathbf{C}_{cond}$ and of changing dielectric polarization
$\mathbf{P}$. The latter, according to the Mossotti hypothesis
(1846) would implicate currents in precisely the same way that
changing charge densities implicate conduction currents. Helmholtz
linked electrodynamic potential $\mathbf{U}$ to the total current
$\mathbf{C}_{tot}$:

\begin{equation}
\mathbf{U}(\mathbf{r})=\int \frac{\mathbf{C}_{tot}(\mathbf{r}^{\prime })}{%
\mathbf{r}_{d}}d^{3}\mathbf{r}^{\prime }+\frac{1-k}{2}\nabla _{\mathbf{r}%
}\int [\mathbf{C}_{tot}(\mathbf{r}^{\prime })\cdot \mathbf{\nabla }_{\mathbf{%
r}^{\prime }}\mathbf{r}_{d}]d^{3}\mathbf{r}^{\prime }  \label{A1}
\end{equation}
where $k$ is undefined constant; $\mathbf{r}$, $\mathbf{r}^{\prime
}$ are
positions of current differential elements and $\mathbf{r}_{d}=\mathbf{r-r}%
^{\prime }$ is the distance between them. This gives the expression for $%
\mathbf{U}$ to which Maxwell's, Weber's and Neumann's theories
lead by setting $k$ equal, respectively, to $0$, $-1$, and $1$.

Dielectric polarization $\mathbf{P}$ in this approach can be
created by electrodynamic potential $\mathbf{U}$ (representing
magnetic forces) as well as by static potential $\Phi $\
(representing electric forces):

\begin{equation}
\mathbf{P}=-{\large \chi }\mathbf{\nabla }\Phi -{\large \chi }A^{2}\frac{%
\partial \mathbf{U}}{\partial t}  \label{A2}
\end{equation}
where the quantity $A$ depends on the medium susceptibility $\chi
$ and, as it will be seen later, defines the propagation speed of
transverse electromagnetic waves. The term $\frac{\partial
\mathbf{U}}{\partial t}$ mathematically describes the interaction
effect of changing currents on charges. Conversely, Helmholtz
assumes that the spacial variation of the static potential $\Phi $
shows the effect of a charge in producing a current (Ohm's law):

\begin{equation}
\mathbf{C}_{cond}=\frac{1}{R}{\large (}A^{2}\frac{\partial \mathbf{U}}{%
\partial t}-\mathbf{\nabla }\Phi {\large )}  \label{A3}
\end{equation}
where $R$ is the resistivity of a conductor.

Helmholtz also considered continuity equation for total currents

\begin{equation}
\mathbf{\nabla }\cdot \mathbf{C}_{tot}+\frac{\partial \rho }{\partial t}%
=0\qquad {\large (}\mathbf{C}_{tot}=\mathbf{C}_{cond}+\frac{\partial \mathbf{%
P}}{\partial t}{\large )}  \label{A4}
\end{equation}
and Poisson's equation for static potential

\begin{equation}
\nabla ^{2}\Phi =-4\pi \rho  \label{A5}
\end{equation}

All these equations (1)-(5) constitute the foundations of
Helmholtz's electrodynamics. In this approach all forces can be
deduced in terms of the potential functions $\mathbf{U}$ and $\Phi
$. In other words it means that the underlying analytical
structure of Helmholtz's electrodynamics can be reduced to the
partial differential equations for both potentials:

\begin{equation}
\nabla ^{2}\mathbf{U}=(1-k)\mathbf{\nabla }\frac{\partial \Phi }{\partial t}%
-4\pi \mathbf{C}_{tot}  \label{A6}
\end{equation}

\begin{equation}
\mathbf{\nabla }\cdot \mathbf{U}=-k\frac{\partial \Phi }{\partial
t} \label{A7}
\end{equation}
Using these equations Helmholtz obtained a general equation for
propagation:

\begin{eqnarray}
-\frac{A}{R}\frac{\partial ^{2}\mathbf{U}}{\partial t^{2}}+\frac{\mathbf{%
\nabla }(\mathbf{\nabla }\cdot \mathbf{U})}{kR}-{\large \chi }A^{2}\frac{%
\partial ^{3}\mathbf{U}}{\partial t^{3}}=\nonumber \\
-\frac{1+4\pi {\large \chi }}{4\pi
{\large \chi }}\mathbf{\nabla }\frac{\partial }{\partial t}(\mathbf{\nabla }%
\cdot \mathbf{U})+\frac{1}{4\pi }\mathbf{\nabla }\times (\mathbf{\nabla }%
\times \frac{\partial \mathbf{U}}{\partial t})  \label{A8}
\end{eqnarray}

If one separates cases for which $\nabla \cdot \mathbf{U}_{t}=0$
(the divergence vanishes for transverse components
$\mathbf{U}_{t}$) and $\nabla
\times \mathbf{U}_{l}=0$ (the curl vanishes for longitudinal components $%
\mathbf{U}_{l}$), the general equation for $\mathbf{U}=\mathbf{U}_{t}+%
\mathbf{U}_{l}$ can be split in two:

\begin{equation}
{\large \chi }\frac{\partial ^{2}\mathbf{U}_{t}}{\partial t^{2}}=\frac{1}{R}%
\frac{\partial \mathbf{U}_{t}}{\partial t}+\frac{1}{4\pi A^{2}}\nabla ^{2}%
\mathbf{U}_{t}  \label{A9}
\end{equation}

\begin{equation}
{\Large \chi }\frac{\partial ^{3}\mathbf{U}_{l}}{\partial t^{3}}=\frac{1}{kR}%
\nabla ^{2}\mathbf{U}_{l}-\frac{1}{R}\frac{\partial ^{2}\mathbf{U}_{l}}{%
\partial t^{2}}+\frac{1+4\pi {\large \chi }}{4\pi kA^{2}}\nabla ^{2}\frac{%
\partial \mathbf{U}_{l}}{\partial t}  \label{A10}
\end{equation}
In non-conducting medium, the resistivity $R$ is infinite and
Helmholtz's equations for propagation result considerably
simplified:

\begin{equation}
\frac{\partial ^{2}\mathbf{U}_{t}}{\partial t^{2}}=\frac{1}{4\pi {\large %
\chi }A^{2}}\nabla ^{2}\mathbf{U}_{t}  \label{A11}
\end{equation}

\begin{equation}
\frac{\partial ^{2}\mathbf{U}_{l}}{\partial t^{2}}=\frac{1+4\pi
{\large \chi }}{4\pi k{\large \chi }A^{2}}\nabla
^{2}\mathbf{U}_{l}  \label{A12}
\end{equation}
Same wave equations can be derived for the scalar potential $\Phi
=\Phi _{t}+\Phi _{l}$.

Originally, Helmholtz's theory was conceived only for polarizable
material medium. In order his equations for propagation to make
sense in empty space, the notion of ether should have been
accepted and the ether itself should have been considered as
polarizable. It was done in an attempt to understand Maxwell's
field equations as a limit case of Helmholtz's theory. This
assumption that the ether exists and has a non-zero susceptibility
$\chi _{0} $ introduces into Helmholtz's scheme \textit{fieldlike}
feature similar in status to the conception of light ether in
Maxwell's approach.

Among researchers in the history of classical electrodynamics
there is a full consensus that to reach formally (i.e. not on a
conceptual level) the Maxwell scheme from Helmholtz's theory
requires two limits: $k=0$ and infinite ether susceptibility $\chi
_{0}$. The former is necessary to obtain Maxwell's equations
(rewritten for $4$-potential) and the latter, to make the velocity
of transverse waves in Helmholtz's theory equal to the velocity of
light $c$. Let us follow both limits, in order to analyze the
limit behavior of transverse and longitudinal waves in Helmholtz's
scheme. First, all basic constants in both theories are to be made
mutually related. The constant $A$ depends on
ether's susceptibility $\chi _{0}$ and can be reorganized into new quantity $%
c^{-1}=A\sqrt{1+4\pi \chi _{0}}$ which is inversely proportional
to the velocity of light $c$ obtained in Maxwell's theory. Hence,
the speed of propagation for transverse waves in Helmholtz's
theory is

\begin{equation}
\mathbf{v}_{t}=c\sqrt{\frac{1+4\pi {\large \chi }_{0}}{4\pi {\large \chi }%
_{0}}}  \label{A13}
\end{equation}

To reach effectively Maxwell's wave velocity $c$, the ether susceptibility $%
\chi _{0}$ must be infinite in Helmholtz's wave equations. In the
case of longitudinal components, the propagation rate with
reorganized $A$ constant is

\begin{equation}
\mathbf{v}_{l}=c\frac{1+4\pi {\large \chi }_{0}}{\sqrt{4\pi k{\large \chi }%
_{0}}}  \label{A14}
\end{equation}

Obviously, the longitudinal wave acquires infinite velocity by
setting $\chi _{0}$ to infinity independently of the value of $k$.
In no way it means that the longitudinal component as a
mathematically valid solution of general Helmholtz's equation for
propagation, lacks any sense or disappears as Hertz assumed in his
analysis of the Maxwell limit of Helmholtz's scheme. The claim
that the limit $k=0$ is necessary to reach the infinite velocity
of propagation came from Helmholtz himself. Perhaps, it was due to
the
erroneous final expression for $\mathbf{v}_{l}=c\sqrt{\frac{1+4\pi \chi _{0}%
}{4\pi k\chi _{0}}}$ given by Helmholtz in his early work (1870),
as commented by Buchwald$^{(14)}$.

In fact, original Maxwell's theory did not predict any
longitudinal waves because basic wave equations were homogeneous
(or sourceless), reflecting the form of continuity equation used
by Maxwell ($\mathbf{\nabla }\cdot \mathbf{C}_{tot}=0$). Later on,
in Lorentz's discussion of Maxwell's theory (1892) a moving charge
as a field source had been incorporated. It marked the main trend
in modification of underlying concepts in original Maxwell's
theory. Primarily, this approach transformed the Maxwell
continuity equation because the conception of source resulted
compatible with Helmholtz's continuity equation. Thus, original
Maxwellian field theory lacked many of the conceptions that many
years later formed part of a unified system known and accepted
nowadays as Lorentz's microscopic theory of electromagnetism.

Lorentz's modification of Maxwell's field approach provided
Lienard (1898) and Wiechert (1901) with inhomogeneous wave
equations which \textit{retarded solutions} were found under the
well-known condition (which was thought to be just experimentally
verified by Hertz (1888)) that electromagnetic interactions
propagate with the velocity of light. It gave a rise to the fact
that longitudinal and transverse components in Lienard-Wiechert
solutions propagate at the same rate. Nevertheless, in many
situations Lienard-Wiechert's longitudinal components can be
eliminated by means of appropriate gauge transformations that
makes their status highly uncertain in the framework of modern
classical electrodynamics. On the contrary, Helmholtz's scheme in
Maxwellian limit offers a more consistent approach to longitudinal
components: longitudinal forces are always present and constitute
bipartite, instantaneous mutual interaction (i.e., they are
irremovable). This scheme, contains electro- and magnetostatics as
naturally valid limit case whereas in the Faraday-Maxwell theory
all static and quasistatic phenomena are totally foreign to the
notion of Faraday's local field.

Another point worth emphasizing here, illustrates the similarity
of Helmholtz's scheme with modern field theories. Helmholtz based
his approach on the notion of electrodynamic and static potentials
(vector and scalar potential function in modern notation).
Potential functions are essential features of all field theories
independently of a particular model of the field. Thus, the novel
character of Helmholtz's electrodynamics consists in the
combination of transverse (local) and longitudinal (non-local)
parts of those potential functions. In modern retrospective, it
means that Helmholtz's approach is purely field scheme, sharing
elements of both local and non-local field theories.

\begin{center}
\textbf{APPENDIX B. The Brief Sequence of Hertz's "Crucial"
Experiments}
\end{center}

From Hertz's published papers we know that he was looking for the
effect of interference of the action that propagated along the
wire with the velocity close to the velocity of light by the
direct action of oscillator that might have very different rate of
propagation. Hertz reasoned in the following way: $(a)$ if both
actions travel at the same rate, the interference picture will be
the same no matter where it is measured (no phase shift observed);
$(b)$ if propagation speeds are not equal, then the phase
difference will depend on the measurement place. In the case, the
direct action of the oscillator has an infinite rate of
propagation, the interference picture should reflect only the
periodicity of the wire action (i.e. the interference will change
sign at every half wave-length of the waves in the wire).

At the end of 1887 Hertz finally undertook this task. Recently
published laboratory notes$^{(22)}$, dairy and letters to parents
can get access to Hertz's first understanding of his results in
such form that did not appear in his printed papers. From
laboratory notes we know that on November 7 Hertz succeeded in
using the resonator to detect the standing electric waves in a
wire connected to one of the oscillators plate. It meant he was
now prepared to undertake interference experiments. Notes dated
November 11-12 record Hertz's intention to construct an experiment
to detect a propagation of the ''direct action'' of the
oscillator. In these first series of experiments Hertz set the
resonator's plane vertically. He could rotate the plane of the
resonator about a vertical axis so that the spark gap of the
resonator could point parallel to the wire (position 1 in Hertz's
terminology especially sensitive to wire action) or parallel to
the oscillator gap (position 2 sensitive to direct action of the
oscillator whether electrostatic or electrodynamic in Helmholtz's
specification).

In his published articles Hertz explained how, rotating the
resonator's plane, can be observed differences in sparking for
different deviations in order to determine the interference
picture in different positions from the oscillator (for greater
details of these experiments see, for instance,
Buchwald$^{(14)}$). On November 11 Hertz succeeded in producing a
detectable interference between the direct action of the
oscillator and the wire waves. The entry of his dairy notes for
November 12 reads: "\textit{Set up experiments on the velocity of
propagation of the electromagnetic effect. Contrary to
expectations, the result is infinite propagation}".

Discouraged by \textit{futile} efforts to detect novel effects
(finite rate of propagation), Hertz put aside experimenting. Later
on for the introduction to "Electric Waves" he gave account about
the first experiment:

\begin{quotation}
"\textit{...Dishearted, I gave up experimenting. Some weeks passed
before I began again. I reflected that it would be quite as
important to find out that electric force was propagated with an
infinite velocity, and that Maxwell's theory was false, as it
would be, on the other hand, to prove that this theory was
correct, provided only that the result arrived at should be
definite and certain.}"
\end{quotation}

Five weeks later, On December 15, Hertz returned to his
experiments again. As it is seen from his dairy, he decided to
write a detailed paper on negative results but only after having
checked and making secure of his previous results. After two days
of laboratory tests, Hertz became convinced that his device was a
very reliable interference detector. This confidence allowed him
to start a new series of experiments performed from December 12
through 21. They were to ensure that the device was properly
constructed and calibrated before setting experiments about the
propagation velocity. After having finished calibrations of his
device, Hertz turned his experimentation to the main goal. On
December 22, he started a next series of experiments for the
resonator's position 1 and 2 (see previous description). Hertz
concluded from his observations that the interference remained in
step with the wire wave. In other words, it confirmed that the
direct action of the oscillator had immeasurably higher speed.

From the letter to his parents written next morning, we get
Hertz's first perception of this result$^{(14)}$:

\begin{quotation}
"\textit{What is the unexpected and to me displeasing result of my
endeavors? The velocity [of the direct action] is not that of
light, but certainly much greater, perhaps infinitely great, at
all events not measurable. Even if it were three times as great,
it could still be measured... Now, there is no arguing with
nature; it must be as it is, but I should have certainly liked it
better to obtain a clear, positive result than this more negative
one... Certainly, caution is indicated here, but once again the
experiments seem all too clear to me.}"
\end{quotation}

On this same day later, December 23, 1887 Hertz put a resonator in
a different orientation. The resonator's plane was now horizontal
(position 3 in Hertz's classification). After Christmas, On
December 26 Hertz began a new series of experiments using the
position 3 with the resonator's gap parallel to the wire. In
Hertz's mind, the resonator was governed entirely by the
oscillator's electrodynamic action (not electrostatic in
Helmholtz's classification). According to the measured
interference phase shift along the wire, Hertz unexpectedly found
that the oscillator's direct action propagated $1.5$ times faster
than the wire wave (i.e. close to the velocity of light). Having
done these promising measurements, Hertz sought a clear
possibility to establish a relation between the speeds of the two
actions: the electrostatic and the electrodynamics in Helmholtz's
classification.

To adapt experiments for this aim Hertz used the position 3 but
with the resonator's gap facing the oscillator. In order to
measure interference only between two actions, Hertz for the first
time removed the wire used for previous interference experiments.
In this configuration the resonator responds to both static and
dynamic actions, though the latter predominates. If the resonator
gap faces the oscillator, the signed value of the total driving
force is a positive contribution of both parts of direct action.
If the gap faces in opposite direction ($180^{0}$ rotation about
the center of the resonator), the contribution of the
electrostatic action will change the sign. By setting the
resonator in different positions Hertz obtained interference marks
which he put in the final table (discussed in the main text
concerning singularities in the near zone). These were Hertz's
first measures of the direct action propagation without the
support of the wire wave of known length. On December 30, Hertz
undertook new observation up to 12 meters.

On January 1, on the first day of 1888 Hertz wrote to his parents
that he was somewhat depressed, anticipating the distasteful risk
of carrying on with additional experiments, "\textit{because one
can only lose by them}". Some days later he began writing up the
article, troubled by doubts that the results might be
"\textit{figments of the imagination}"$^{(14)}$. Finally, Hertz
calmed his doubts and by the end of January wrote the paper
entitled, "\textit{On the Finite Velocity of Propagation of
Electromagnetic Action}".

The history of physics researchers seemed to show no interest to
the fact that Hetrz's drastic change in attitude towards his
previous "negative" results was not supported by additional
reliable experimental measurements. The confirmation of this
change is clearly seen in Hertz's Introduction to
"\textit{Electric Waves}" where he was forced to admit many doubts
on the reliability and accuracy of his first \textit{negative
experiments}.

Nevertheless, in modern retrospective the second series of Hertz's
experiments and the table obtained on December 22 (where he
confirmed his previous "negative" results) do not seem to support
very strongly Hertz's allegations on inaccuracy. This is also
Buchwald's opinion in his own words$^{(14)}$:

\begin{quotation}
\textit{...Neither the 3.4 we find in Hertz's undiscussed average
not the 2.9 obtained directly from his table easily sustains a
positive argument for finite propagation. On the contrary, the 2.9
result falls right between the 2.66 and 3.1 values for the wire
wave's half-length that he previously found, and the 3.4 result
barely misses the first, and includes the second, value if we
allow the inaccuracy to be no larger than the standard deviation
among the measurements. If, as Hertz would certainly have had to
admit, the inaccuracies are in fact somewhat larger than this,
then his numbers cannot be instrumentally distinguished from one
another. These results, as they stand, accordingly nicely sustain
Hertz's anticipated negative outcome. But because Hertz later
decided his experiments had positive results, in his published
papers he had to argue the apparent implication of the table away,
which he did in two ways: first, by saying that the alternation is
not at all precise (which might, however, speak just as well to
the experiment's accuracy as to its implications) and, second, by
capitalizing on what might otherwise appear to be inaccuracy in
observation, namely, the frequent appearance of four or more zero
points together at distances past 3 m ...}
\end{quotation}

Thus, it is not still clear why Hertz did not overcome
"\textit{the distasteful risk of carrying on with additional
experiments}" and did not repeat his first experiments in order to
remove or confirm his doubts in their inaccuracy. His published
papers, laboratory notes and dairy, unfortunately, provide little
help here.

{\large {\bf REFERENCES}}

\begin{enumerate}

\item{} S. Laplace, \textit{Mecanique Celeste}, Book X, 22, (Paris, 1799)

\item{} T. Van Flandern and J.-P. Vigier, \textit{Foundations of Physics},
\textbf{32}(7), 1031-1068 (2002)

\item{} I. Newton, \textit{Third Letter to Bentley}, \ from Work of
Richard Bentley, III, 211

\item{} S. Laplace, \textit{Exposition du System du Monde}, Book IV
(Paris, 1796)

\item{} P. Graneau and N. Graneau, \textit{Newton versus Einstein:
How Matter Interacts with Matter} (Carlton Press, New York, 1993)

\item{} D. Hoffmann, "Heinrich Hertz and the Berlin School of
Physics", in \textit{Heinrich Hertz: Classical Physicist, Modern
Philosopher} , eds. D. Baird, R.I.G. Hughes and A. Nordmann,
(Kluwer Academic, Dordrecht, 1998), 1-8

\item{} M. Heidelberger, "From Helmholtz's Philosophy of Science
to Hertz's Picture-Theory", in \textit{Heinrich Hertz: Classical
Physicist, Modern Philosopher} , eds. D. Baird, R.I.G. Hughes and
A. Nordmann, (Kluwer Academic, Dordrecht, 1998), 9-24

\item{} J. Maxwell, \textit{On Faraday's Line of Force},
Scientific Papers, Vol.\textbf{1}, 160 (1864)

\item{} M. Hesse, "Action at a Distance in Classical
Physics" \textit{ISIS} \textbf{46, }337-353 (1955)

\item{} H. Hertz, "On the Fundamental Equations of Electromagnetics
for Bodies at Rest", in \textit{Electric Waves}, Collection of
Scientific Papers (Dover, New York, 1962), 195

\item{} M. Hesse, \textit{Forces and Fields: The Concept of
Action at a Distance in the History of Physics} (Thomas Nelson and
Sons Ltd., London, 1961)

\item{} H. Helmholtz, \textit{Wissenschaftliche Abhandlungen}, Vol. \textbf{1},
(Barth, 1882), 611-628

\item{} P.S. Kudryavtzev, \textit{History of Physics}, vol. \textbf{2}, (Moscow
University Press, Moscow, 1956), 206-213 (in Russian)

\item{} J. Buchwald, \textit{The Creation of Scientific Effects: Heinrich
Hertz and Electric Waves} (The University of Chicago Press,
Chicago, 1994)

\item{} A.E. Woodruff, "The Contribution of Hermann von
Helmholtz to Electrodynamics" \textit{ISIS} \textbf{59}, 300-311
(1968)

\item{} H. Helmholtz, \textit{Wissenschaftliche Abhanlugen}, vol.\textbf{1} (Barth,
1882), 556

\item{} H. Hertz, "On the Finite Velocity of Propagation of
Electromagnetic Actions" in \textit{Electric Waves}, 110 (1888)

\item{} ibid., "On Very Rapid Electrical Oscillations"
in \textit{Electric Waves}, 29-53 (1887)

\item{} ibid., "On the Finite Velocity of Propagation of
Electromagnetic Actions" in \textit{Electric Waves}, 108 (1888)

\item{} ibid., 121

\item{} J. Cazenobe, \textit{Archives Internationale
d'Histoire des Sciences}, \textbf{32}, 236-265 (1982)

\item{} M. Doncel, "Heinrich Hertz's Laboratory Notes
of 1887" \textit{Archive for History of Exact Sciences},
\textbf{49}, 197-270 (1995)

\item{} M. Doncel, "On Hertz's Conceptual Conversion:
From Wire Waves to Air Waves" in \textit{Heinrich Hertz: Classical
Physicist, Modern Philosopher}, eds. D. Baird, R.I.G. Hughes and
A. Nordmann (Kluwer Academic, Dordrecht, 1998), 73-87

\item{} J. Z. Buchwald, "Electrodynamics in Context: Object States,
Laboratory Practice and Anti-Romanticism" in \textit{Hermann von
Helmholtz and the Foundations of Nineteenth-Century Science},
Edited by D. Caham (University of California Press, Berkely,
1993), 345-368

\item{} H. Hertz, \textit{Electric Waves}, Collection of Scientific Papers,
(Dover, New York, 1962), 15

\item{} ibid., "The Forces of Electric Oscillations, Treated
According to Maxwell's Theory" in \textit{Electric Waves}, 137
(1889)

\item{} ibid., 151-152

\item{} W. Panofsky and M. Phillips, \textit{Classical Electrodynamics and
Magnetism} (Addison Wesley P.C., Massachusetts, Second Edition,
1962), 259-260

\item{} H. Hertz, "The Forces of Electric Oscillations, Treated
According to Maxwell's Theory" in \textit{Electric Waves}, 149
(1889)

\item{} ibid., "On the Mechanical Action of Electric Waves in
Wires" in \textit{Electric Waves}, 187 (1891)

\item{} M. Heidelberger, "From Helmholtz's Philosophy of Science
to Hertz's Picture-Theory" in \textit{Heinrich Hertz: Classical
Physicist, Modern Philosopher} , eds. D. Baird, R.I.G. Hughes and
A. Nordmann (Kluwer Academic, Dordrecht, 1998), 18

\item{} H. Hertz, \textit{Electric Waves}, Introduction, 18-19

\item{} P. Duhem, \textit{The Aim and Structure of Physical
Theory} (Princeton University Press, Princeton, 1954)

\item{} P. Duhem, \textit{Les Theories Electriques de J. Clerk
Maxwell}, Paris, 1902 (quoted from A. O'Rahilly,
\textit{Electromagnetic Theory: A Critical Examination of
Fundamentals}, vol. 1 (Dover, New York, 1965), 161-180)

\item{} R.P. Feynman, \textit{Lectures on Physics: Mainly
Electromagnetism and Matter} (Addison-Wesley, 1964)

\item{} A. Chubykalo and R. Smirnov-Rueda,
\textit{Physical Review E}, \textbf{53}, 5373-5381 (1996)

\item{} A. Chubykalo and R. Smirnov-Rueda,
\textit{Modern Physics Letters A}, \textbf{12}(1), 1-24 (1997)

\item{} A. Einstein, \textit{The Meaning of Relativity}, (Princeton
University Press, Princeton, 5th Edition, 1955)

\end{enumerate}

\end{document}